\documentclass[%
 aip,
 amsmath,amssymb,
preprint
]{revtex4-1}

\usepackage{color}
\usepackage[usenames,dvipsnames]{xcolor}
\usepackage{graphicx}
\usepackage{dcolumn}
\usepackage{bm}
\usepackage{float}
\usepackage[utf8]{inputenc}
\usepackage[T1]{fontenc}
\usepackage{mathptmx}
\usepackage{hyperref}
\hypersetup{
    colorlinks=true,
    linkcolor=blue,
    urlcolor=blue,
    citecolor=blue,
}
\begin{document}

\title{High-Speed Ultrasound Imaging of Laser-Induced Cavitation Bubbles}


\author{S. Izak Ghasemian}%
\email{saber.izak@ovgu.de}
\affiliation{Institute of Physics, Otto-von-Guericke Universität, Universitätsplatz 2, 39016 Magdeburg, Germany}
\author{F. Reuter}%
\affiliation{Institute of Physics, Otto-von-Guericke Universität, Universitätsplatz 2, 39016 Magdeburg, Germany}
\author{C.D. Ohl}%
\affiliation{Institute of Physics, Otto-von-Guericke Universität, Universitätsplatz 2, 39016 Magdeburg, Germany}
\affiliation{Research Campus STIMULATE, University of Magdeburg, Otto-Hahn-Straße 2, 39106 Magdeburg, Germany.}

\date{\today}

\begin{abstract}
While the ultrasound and cavitation based therapies have mushroomed over the years, there is a lack of online monitoring the cavitation bubble dynamics in biological tissue. {\color{black}Here we demonstrate that with ultrasonic plane wave imaging the fast dynamics of single cavitation bubbles can be resolved non-invasively in a tissue mimicking material}. Due to the high contrast of the bubbles, plane wave compounding is not necessary and frame rates of up to 74 kHz can been achieved with a research grade ultrasound scanner. Comparison with simultaneous high-speed imaging demonstrates excellent agreement of both measurement modalities in water and in a tissue mimicking material.
\end{abstract}

\maketitle

Deposition of energy within liquids can result in the formation of voids that expand and collapse. This phenomenon is termed cavitation and has applications in many biomedical sciences. Cavitation based therapies are playing an increasingly important role in medical treatment such as opening of the blood-brain barrier for drug delivery into the brain, kidney lithotripsy for the comminution of kidney stones and histotripsy for liquefying benign and malign tissues\cite{Hu2020, Ramaswamy2015}. While in these applications cavitation bubbles are utilized, their control through exposure parameters to improve precision demands for a monitoring of the bubble activity. Conventionally this is done by looking at the echogenicity of the tissue with B-mode ultrasound. While this offers a look at residual bubbles it does not allow to follow the bubble dynamics, i.e. their maximum radius, after bounces, or their motion. This lack of imaging modality stimulated our research to record individual bubbles oscillating in a tissue mimicking material (TMM). Cavitation bubbles can be generated in TMM using different techniques, here we chose an optic cavitation\cite{lauterborn2010physics} approach with a pulsed and focused laser. These bubbles are created by forcing a dielectric breakdown in the transparent material thereby generating a short lived plasma that after recombination is the nucleus for an explosively expanding vapor bubble. 
Dynamics of bubbles in water or transparent TMMs is traditionally captured with optical methods. Due to the limited transparency, optical recording of bubbles in a real biological tissue raises challenges with respect to a suitable wavelength and depth of imaging. In many situation it is even impossible to obtain optical access\cite{zhan2014optical}. 
{\color{black}Here we aim towards overcoming that problem showing the feasibility with an acoustic technique employing a commercial ultrasound imaging transducer in a tissue mimicking material. }
 {\color{black} The use of the tissue mimicking material allows for the generation of single, controlled cavitation bubbles with optic cavitation as well as their optic high speed imaging for comparison.}

Previous work has addressed the task of measuring cavitation bubbles dynamics by acoustic means through passive and active techniques. Karpiouk et al.\cite{Karpiouk2007} have shown that the dynamics of bubbles can be captured precisely using a single element ultrasound transducer. They generated a laser induced cavitation bubbles in water and assumed that with unchanged laser energy, the dynamics of the bubble is repeatable. The bubble radius was measured from the transit time of an acoustic burst at a frequency of 25\,MHz and reflect off the cavitation bubble. Then, by repeated runs at variable delays between the laser pulse and the transducer pulse, the overall bubble dynamics was reconstructed from the collected transit times. 
They obtained good agreement with the expected radial dynamics from a Rayleigh-type bubble model for the studied bubbles that had maximum radii of about 700\,$\mu$m and 35\,$\mu$m, respectively \cite{Karpiouk2007, Aglyamova2008}. While this demonstrated the possibility to measure with sufficient spatial resolution the radial dynamics of a cavitation bubble, the repeatability that is assumed in these experiments is generally in medical applications not given. Also, the use of a single element transducer is only suitable for spatially localized cavitation. While there may be situations such as in eye surgery\cite{Juhasz2000} but for less predictable and as well as for non-repeatable bubble dynamics high-speed acoustic imaging is deemed necessary. 
The challenge to capture the dynamics of transient cavitation bubbles is their short lifetime and fast dynamics. For example, a cavitation bubble with a maximum radius of $500\,\mu$m has a lifetime in a liquid of $T_\textrm{L}=2\,T_\textrm{c}\approx (\rho/p)^{1/2}\approx 100\,\mu$s, where $\rho$ is the density and $p$ is the static pressure, and $T_\textrm{c}$ is the Rayleigh collapse time. Thus to record in water at standard pressure two images of the bubble dynamics demands for a frame rate of 20000 frames/s. \\
Plane wave imaging\cite{Tanter2015} is an established acoustic technique that has been introduced to overcome conventional beam-forming based techniques\cite{Szabo2004} that typically operate at 20-50 frames/s in B-mode. As there is no beam focusing done in this technique, the quality of reconstructed images from single plane waves are generally too poor for most applications and compounding techniques have been developed. There the waves are focused in the medium virtually by sending the plane waves under different angles and by digitally processing the received signals. Montaldo et al.\cite{Montaldo2009} have shown that plane wave compounding with a sufficient number of tilted angles may even achieve the quality of an optimum multifocus B-mode image in terms of signal-to-noise ratio and contrast. However and as always in life, there is a trade-off. Here the trade-off is between the quality of the ultrasound images and the imaging frame rate which reduces with more angles. {\color{black} In general, frame rate in plane wave imaging is a function of imaging depth, speed of sound in the imaging medium, number of compounded waves and also hardware limitations for switch between transmit and receive mode.} In this work we utilize plane wave imaging to capture the dynamics of a transient cavitation bubbles with high temporal and spatial resolution. The acoustically measured motion of the bubble as a function of time is benchmarked against a simultaneously conducted optical high-speed imaging. 


Before we utilize plane wave imaging for the measurement of cavitation bubble dynamics, we study the image quality of compounding plane waves to image bubbles in a test experiment. This is important as wave compounding reduces the effective frame rate and if possible should be limited or even avoided.
Imaging is done with a research ultrasound system (Vantage 64LE; Verasonics, Kirkland, WA, USA, sampling rate 62.5\,MHz) connected to a 5-MHz linear probe (ATL L7–4; Philips, Amsterdam, The Netherlands).

\begin{figure}
\includegraphics[width=\textwidth]{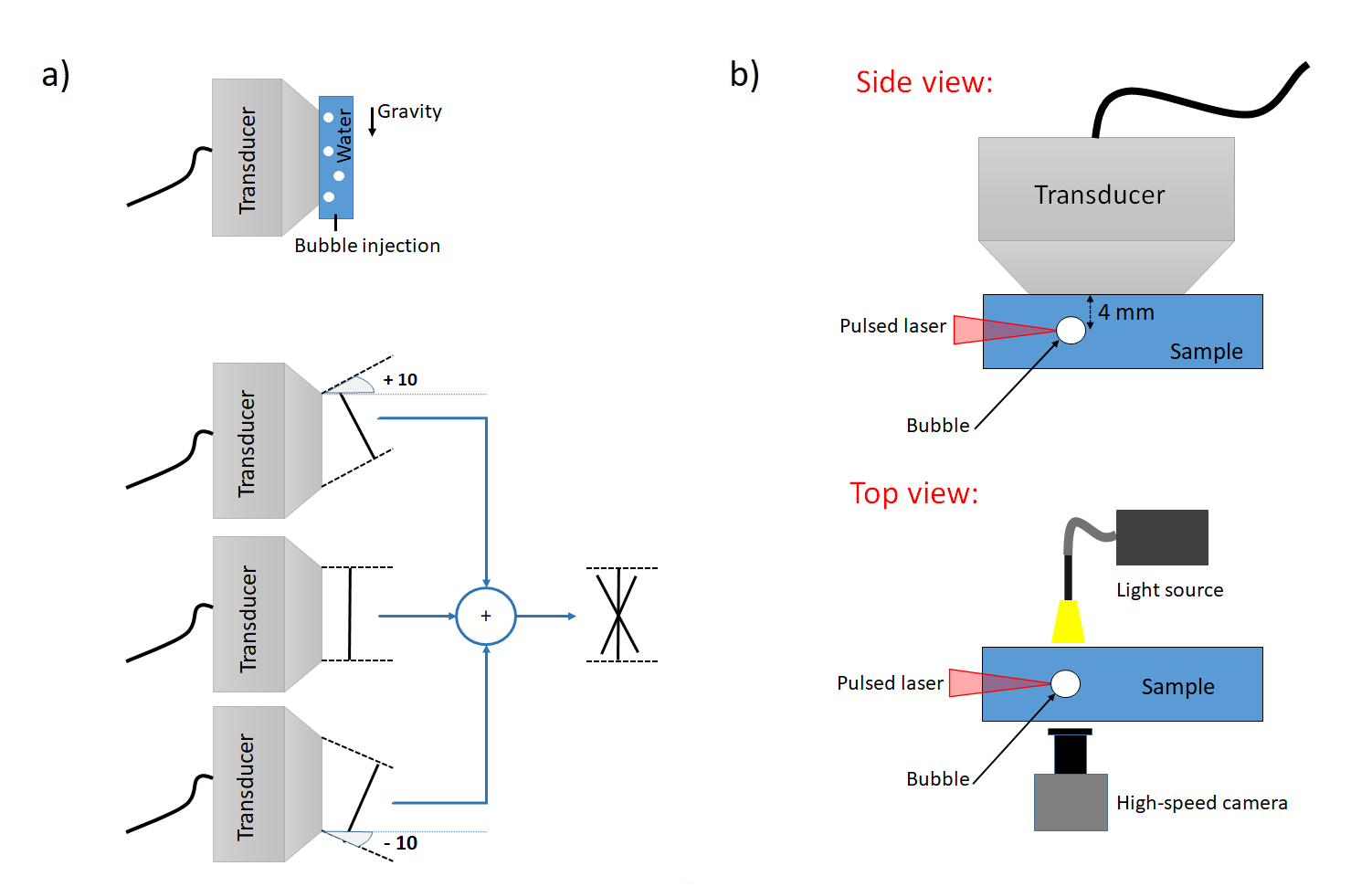}
\caption{Experimental setup for: a) Top: plane wave imaging of rising bubbles in water {\color{black}with an imaging depth of $12\,$mm.} Bottom: schematic of plane wave compounding with three different tilted angles. b) Optical high speed imaging and ultrasound imaging of laser-induced cavitation bubbles in water and gelatin {\color{black}with an imaging depth of $6\,$mm. The bubble is generated at a distance of about $4\,$mm below the transducer.}}
\label{setupSketch}
\end{figure}

Gas bubbles are generated by injecting air through a submerged needle and are then rising freely and rather slowly over a distance of about $5\,$cm to the surface of a container, see Fig.~\ref{setupSketch}a. The home built cuvette has an acoustically transparent window on one side where the ultrasound imaging probe is attached to and glass windows on the other face sides for optical imaging from the side.
 
\begin{figure}
\includegraphics[width=\textwidth]{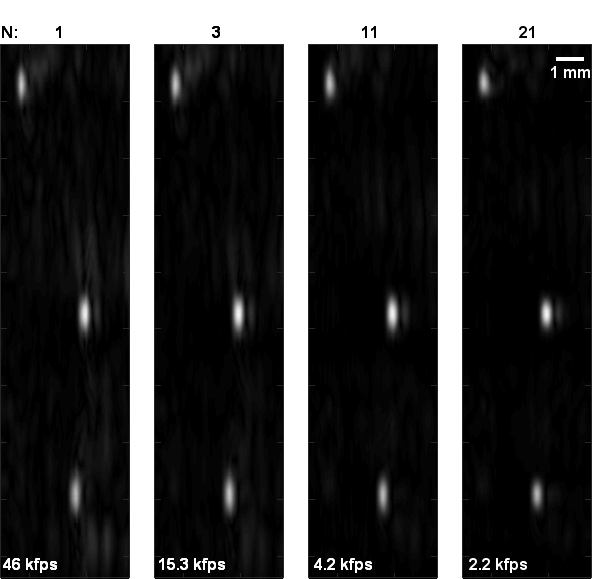}
\caption{Qualitative comparison of single plane wave and plane wave compounding imaging of a bubble column in water with different number of tilted angles. $N$ denotes the number of plane wave compounding. (Multimedia view)}
\label{compoundingComparison}
\end{figure}

Figure~\ref{compoundingComparison} compares the images obtained with an increasing number $N$ of compound waves, reconstructed for a depth of $3\,$mm to $7.5\,$mm. The experiment was conducted in a single run with 21 different tilt angles from $-10^\circ$ to $+10^\circ$ but in the first three images only $N=1, 3, 11$, plane waves are used to process the compound image. For $N=1$, the single plane wave at an angle of $0^\circ$ is chosen, for $N=3$, compound waves with angles $0^\circ$ and $\pm 10^\circ$ are chosen, and for $N=11$, and $N=21$ intermediate angles are additionally considered. The millimeter-sized bubbles rise at a velocity of 0.3 m/s, thus move only 6 $\mu$m between two compound waves, {\color{black}and even for highest number of compounding less than half of the acoustic wavelength, thus we assume them as static bubbles here}.

Due to the strong acoustic impedance ratio between water and air, a strong acoustic reflection is caused at the transducer facing, liquid-gas interface of the bubbles. Their vertical size reveals the approximate bubble diameter, while the horizontal size is a result of the finite duration of the acoustic probing pulse.
The reconstructed movie of these rising bubbles is available as supplementary material (Multimedia view). Interestingly, already with a single plane wave, the ultrasound image provides sufficient contrast for measuring the location and extent of the bubbles. With more angles, the signal to noise ratio increases, imaging artefacts reduce, and the vertical length reduces. Yet, for detecting bubbles and measure their locations, we do not depend on compounding. This simplifies processing and benefits the achievable frame rate for the study of rapidly expanding and collapsing cavitation bubbles.

Next we study the fast dynamics of expanding and collapsing bubbles with single plane waves. A sketch of the experimental setup is shown in Fig.~\ref{setupSketch}b. The geometry is chosen, that the acoustic and optical path are perpendicular to each other. The bubble is viewed from the top with the ultrasound probe and from the side with the optical high-speed camera. Single, laser induced cavitation bubbles are created by focusing a collimated laser pulse (Litron Nano series, Q-switched Nd:YAG, 6\,ns, wavelength 1064\,nm) into deionized water with a microscope objective (Leitz, Wetzlar UM 20/0.33).
At the focal region of the lens, the bubble is generated through an optical breakdown. The dynamics of the bubble (expansion, collapse, and rebound) is recorded with a high speed camera (Photron, FASTCAM Mini, AX200) equipped with a macro lens (Canon MP-E 65\,mm f/2.8 $1-5\times \text{Macro}$) simultaneously with the ultrasound imaging system. The ultrasound imager is recording the radio frequency (RF) data from single plane waves. The timing of the devices is controlled with a digital delay generator (BNC 525, Berkeley Nucleonics) such that ultrasound imaging and the camera are triggered simultaneously with the start of the laser pulse. 

The same experimental setup is used for recording the bubble dynamics in gelatin that has a very similar acoustic impedance as water. Gelatin is a hydrogel that is commonly used for TMM\cite{McGarry2020}. Its elastic properties can be adjusted by changing its concentration. Several recent works have utilized gelatin to study cavitation bubble dynamics\cite{Kang2018, Oguri2018, Hopfes2019} in an elastic material. Here, the samples are prepared from powdered gelatin (Gelatin 250 bloom, Yasin Gelatin CO., LTD). It is mixed with deionized water at a mass ratio of 4\% (gelatin to water) and dissolved in a flask on a hot plate with a magnetic stirrer similar to a previously described procedure\cite{Rapet2019}. The hot mixture is poured into a home build cuvette (dimensions of $35\times25\times55\,$mm$^3$) which is open from top for ultrasound imaging and optically transparent from three sides for accurate laser focusing, clear observation and illumination, see Fig.~\ref{setupSketch}b. The samples, after cooling down to room temperature, are stored in a fridge. Before use, we ensure that the samples have reached room temperature again. To reduce ultrasound reflection, the bottom wall of cuvette is consists of a low density polyethylene film and additionally the cuvette is mounted on top of a water tank with depth of 60\,mm.

Both in water and in gelatin, the cavitation bubble expands explosively first to a maximum radius before it collapses and then runs through a number of rebounds. After some time, the bubble oscillations have ceased in contrast to water a long-term stable gas bubble remains in gelatin. We want to note that while in gelatin a stable bubble after the cavitation events remains near the laser focus, in water the bubble collapse is more violent resulting into multiple bubble fragments. Then the tiny fragments either float due to buoyancy and their non-condensable gas content diffuses back into the liquid causing their dissolution. Bubbles generated in gelatin over long time (not shown here) either shrink due to elastic forces from the gelatin or they continue to grow by diffusion if the gelatin is sufficiently supersaturated gelatin\cite{Ando2019}.

\begin{figure}
\includegraphics[width=0.8\textwidth]{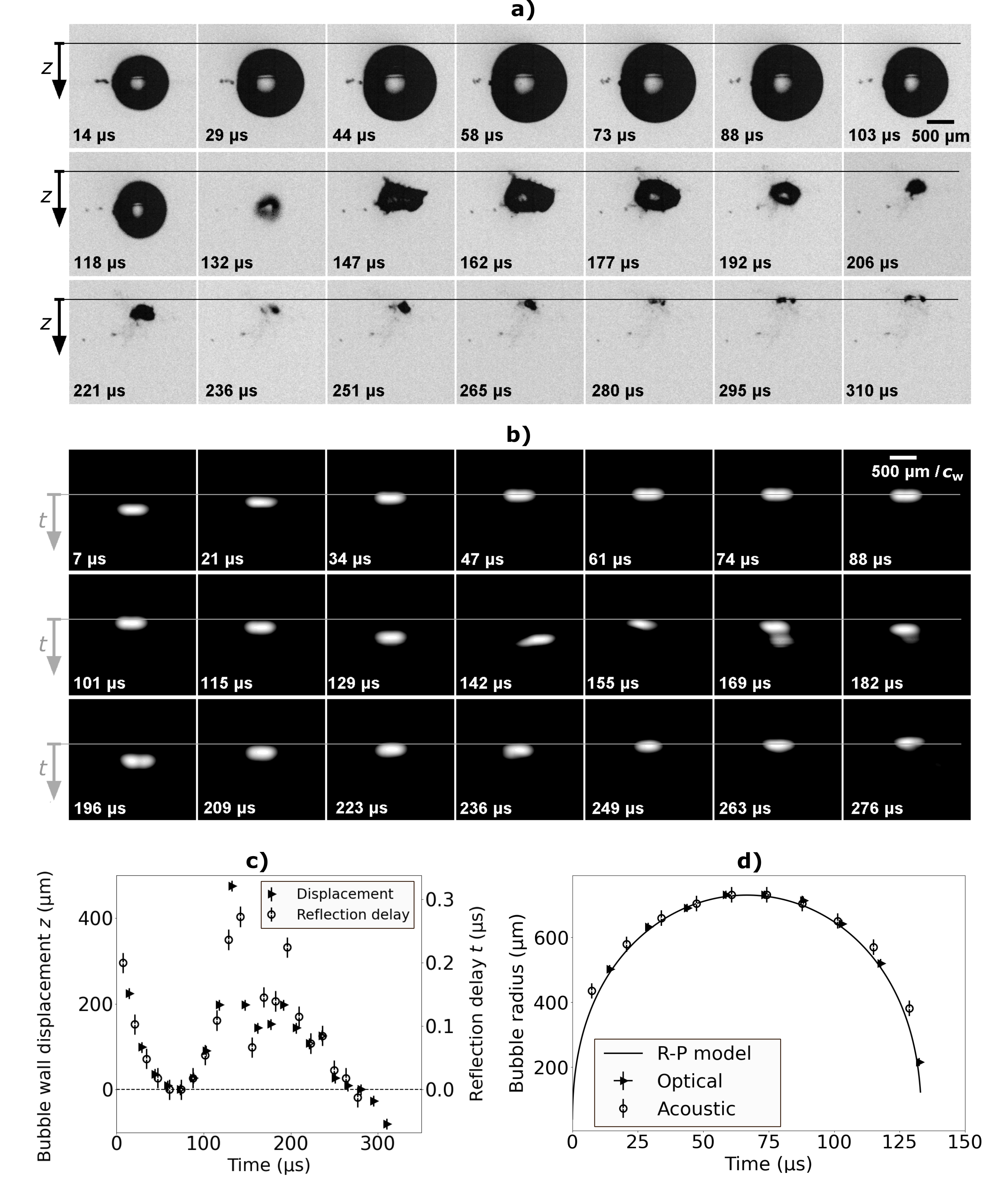}
\caption{a) Selected frames of the dynamics of a laser induced cavitation bubble nucleated in water recorded with the high-speed camera at an exposure time of $4\,\mu$s. b) Reconstructed ultrasound images of the same bubble {\color{black} with single plane wave imaging ($N=1$)}. The ultrasound transducer is aligned horizontally on the top of each frame. The images in a) and b) share the same geometric scale. Assuming a constant speed of sound of $c_{\textrm{w}}=1480\,$m/s, {\color{black} a time axis is defined in ultrasound images.} c) In the left axis: Displacement of upper bubble wall with respect to its position at maximum size, extracted from high speed camera, and in the right axis: Time delay between the reflection of ultrasound wave from bubble upper wall in each frame and the frame of maximum bubble. d) Bubble radius as a function of time from high-speed camera, ultrasound device and Rayleigh-Plesset model.}
\label{waterData}
\end{figure}

Fig.~\ref{waterData}a shows selected frames from an optical high-speed recording of the cavitation bubble during expansion, collapse and rebound in water. The bubble is generated at a distance of about 5\,mm below the ultrasound head. It reaches a maximum radius of $R_\textrm{max}=730\,\mu$m and collapses 130 $\mu$s after generation. We plot additionally a line representing the origin of the $z$-axis pointing downwards. The position $z=0$ is where the upper bubble interface obtains its maximum extend during the first expansion. For the frames before the first collapse, when the bubble center is almost stationary and the bubble mostly spherical, the difference in $z$ between two sequential frames is approximately the change in radius. After collapse however, the bubble center moves and the bubble translates towards the transducer that acts as an attractive rigid boundary. From here on the displacement of $z$ does not represent the change in bubble radius but includes also the translation of the bubble.

Ultrasound images are shown in Fig.~\ref{waterData}b. They are constructed from the RF data with the Delay Multiply And Sum (DMAS) beamforming algorithm implemented in the ultrasound toolbox USTB from \cite{Rodriguez-Molares2017}. {\color{black}The bright spots in the reconstructed ultrasound images indicate the position of highest reflection which is the bubble boundary located closest to the transducer. This region is imaged and thus a deduction of the actual bubble size is not directly possible from the acoustic image. Although we didn't study minimum bubble size that can be resolved, at $276\,\mu$s, when the dimension of the bubble is only $60\,\mu$m in vertical and $300\,\mu$m in horizontal direction, it can be well imaged by the acoustic technique.}
Instead of a $z$-axis, we define a time axis with a positive direction downward. The time $t=0$ is conveniently set to the time of the shortest reflection delay during the first oscillation cycle, i.e. at maximum bubble expansion. The bright spots are caused by the reflection of the ultrasound at the bubble surface facing the ultrasound transducer. Assuming an ideal soft reflector, a position stable bubble, and a constant speed of sound in the liquid, the following relation holds between reflection delay and instantaneous bubble radius $R(t)$: $t\propto R_\textrm{max}-R(t)$, with the inverse speed of sound, $c^{-1}$, being the proportionality coefficient. Yet, after the first collapse the movement of bubble centroid and the non-spherical shape does not provide such a simple relationship.\\
To read the reflection delay $t$ from the acoustic images, a 2D Gaussian is fitted on the intensity planes of Fig.~\ref{waterData}b, and the maximum of the Gaussian fit is taken as the position of the bubble wall located closest to the transducer.
The position of the upper bubble wall $z$ and the reflection delay $t$ are plotted in Fig.~\ref{waterData}c as a function of time. The left vertical axis depicts the displacement of the bubble upper wall from its position at maximum size. The acoustic reflection delay, in microseconds is shown on the right vertical axis. Both axes share the same origin. 
Assuming no motion of the bubble center for the first oscillation cycle, the bubble radius evolution extracted from optical and ultrasound images is shown in Fig.~\ref{waterData}d and compared with the Rayleigh-Plesset model. They are all in a good agreement. The radius from the optical camera has been measured from the images directly. However, in the ultrasound images, as we can only see the upper wall of bubble and not the bubble center, extracting the absolute value for the bubble radius in each frame is not possible and we can only measure the change in bubble radius. To have the absolute bubble radius with time, we need to localize the center of bubble or obtain an absolute bubble radius through other means. To overcome this ambiguity we use the maximum bubble radius from the high speed camera to find absolute values for the bubble radius as a function of time in the ultrasound images. It remains to be shown that the center of the bubble can be determined from the spherical expanding shockwave that is emitted during bubble generation and recorded with the ultrasound probe. The origin of this wave and thus the center of the bubble may be obtained by back-tracking the recorded pressure pulse.

\begin{figure}
\includegraphics[width=\textwidth]{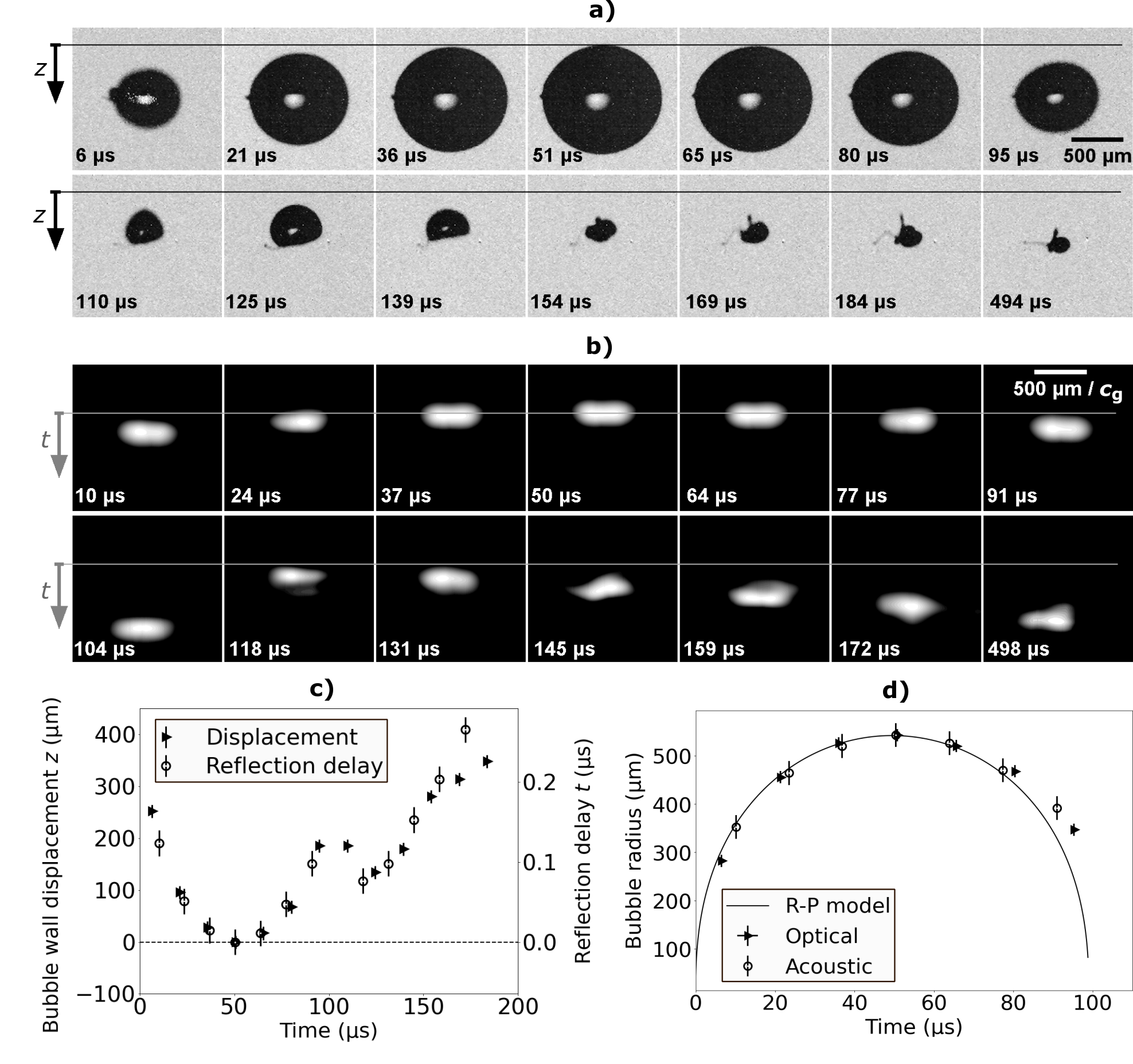}
\caption{a) Selected frames of the dynamics of a laser induced cavitation bubble nucleated in gelatin recorded with the high-speed camera at an exposure time of $4\,\mu$s. b) Reconstructed ultrasound images of the same bubble {\color{black} with single plane wave imaging ($N=1$)}. The ultrasound transducer is aligned horizontally on the top of each frame. The images in a) and b) share the same geometric scale, assuming a constant speed of sound of $1540\,$m/s, {\color{black}a time axis is defined in ultrasound images.} c) In the left axis: Displacement of upper bubble wall respecting its position in its maximum size extracted from high speed camera, and in the right axis: Time delay between the reflection of ultrasound wave from bubble upper wall in each frame and the frame of maximum bubble. d) Bubble radius as a function of time from optical high-speed imaging, acoustic high-speed imaging, and the Rayleigh-Plesset model.}
\label{gelatinData}
\end{figure}

The same experimental setup has been used to measure bubble radius in gelatin. The difference here is that although the center of bubble moves after collapse, the bubble does not rise to the surface, but after oscillations ends up as a stably entrapped gas pocket in the gelatin.
In Figure~\ref{gelatinData} the results are present the same way as in the previous figure. The generated bubble here has a maximum radius of 540\,$\mu$m with the life time of about $T_\textrm{L}\approx104\,\mu$s. In the last optical image the remaining gas bubble can be seen. In Fig.~\ref{gelatinData}b there is one outlier in the data, which we would like to highlight, too. At $t=104\,\mu$s the shock wave emitted from the bubble collapse is acoustically imaged and not the reflection on the bubble interface.
The amplitude of the shock wave is considerably stronger than the reflected signal with the result of a erroneous reconstruction of the bubble.
It can be seen that in the first bubble cycle, the acoustic and camera data match favourably. However, with the Rayleigh-Plesset model the bubble life time is shorter. Considering the elasticity of the gelatin~\cite{Manica2021} is likely to result in a better fit for the collapse.


We have presented an ultrasonic plane wave imaging techniques that is able to capture the dynamics of sub-millimeter sized cavitation bubbles at frame rates of up to 74.2 kHz. Data taken with this technique was validated with optical high speed imaging and the Rayleigh-Plesset model of bubble oscillation in {\color{black} water and a tissue mimicking material. To be able to generate controlled cavitation bubble and also to have optical imaging for comparison, we used a transparent material as TMM.}
Although the acoustic wavelength is only about $300\,\mu$m the high acoustic contrast and the sufficiently fast sampling rate allows measuring the reflection delay with an accuracy comparable to that of the high-speed camera. It is important to note that already single plane wave imaging obtains sufficiently high quality data, thus high acoustic framing rates can be achieved. The results in a tissue mimicking gelatin material are of similar quality as in water and thus we suggest that plane wave imaging is a viable solution to monitor transient cavitation in non-transparent tissue{\color{black}, still in vivo experiments are needed.} 
Further, the bubble dynamics in tissue depends on its elastic properties~\cite{Manica2021} and using plane waves monitoring the progress of cavitation based therapies in tissues becomes feasible. Also the monitoring of the fast bubble dynamics in thermal based therapies of tissues such as radio wave ablation could benefit from high-speed plane wave imaging and its reconstruction of the bubble dynamics. {\color{black}We showed the feasibility of the high speed plane wave imaging technique in 2D, an extension to 3D imaging with appropriate 3D transducers should be straight forward.} A remaining challenge is the extension of the technique to monitor deeper into the tissue while not reducing the framing rate. This may be possible be keeping a high pulse rate excitation combined with an analysis of the receive signals for specific delay times (gates) similar to the well established high-pulse repetition frequency Doppler velocimetry\cite{Szabo2004}.

\section*{Acknowledgement}

This project has received funding from the European Union’s Horizon 2020 research and innovation programme under the Marie Skłodowska-Curie grant agreement No 813766. We dedicate this work to Xiaoming Ji.

\section*{Data Availability}

The data that support the findings of this study are available from the corresponding author upon reasonable request.

\bibliography{paper}
\newpage

\end{document}